\documentclass{article}
\usepackage{amsmath,amssymb}
\usepackage{nccmath,xfrac}
\usepackage[margin=1in]{geometry}
\usepackage{graphicx}
\usepackage{hyperref}
\usepackage{mathtools}
\usepackage[final]{changes}
\usepackage{color}
\usepackage{enumitem}
\usepackage{authblk}
\usepackage{comment}

\title{Martingale properties of entropy production and a generalized
  work theorem with decoupled forward and backward processes}
\author[1]{Xiangting Li} \author[1,2]{Tom Chou} \affil[1]{Department
  of Computational Medicine, University of California, Los Angeles, CA
  90095} \affil[2]{Department of Mathematics, University of
  California, Los Angeles, CA 90095}

\date{\today}
\begin{document}
\maketitle
\begin{center}
    {\bf \large Abstract}
\end{center}

By decoupling forward and backward stochastic trajectories, we
construct a family of martingales and work theorems for both
overdamped and underdamped Langevin dynamics. Our results are made
possible by an alternative derivation of work theorems that uses tools
rom stochastic calculus instead of path-integration. We further
strengthen the equality in work theorems by evaluating expectations
conditioned on an arbitrary initial state value. These generalizations
extend the applicability of work theorems and offer new
interpretations of entropy production in stochastic systems. Lastly,
we discuss the violation of work theorems in far-from-equilibrium
systems.

\vspace{1em}
\section{Background and Introduction}
\begin{figure}
  \centering
  \includegraphics[width = 6.4in]{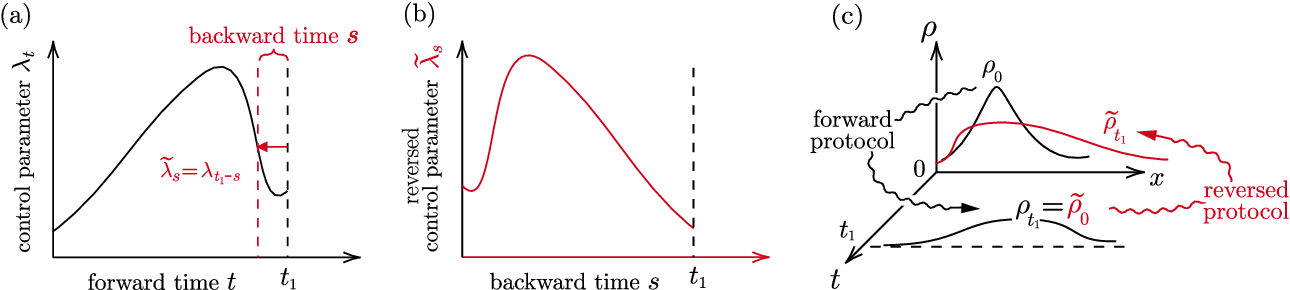}
  \caption{Schematic of \added{standard backward driving protocols and
      the associated probability densities.} (a) The forward driving
    protocol $\lambda_t$ as a function of forward time $t$. (b) The
    backward driving protocol $\tilde{\lambda}_{s}$ is obtained by
    counting time backwards from the terminal time $t_1$. (c)
    Probability densities of the forward ($\rho$) and the standard
    backward ($\tilde{\rho}$) processes described in previous work
    \cite{Manzano2021February}. With an initial distribution
    $\tilde{\rho}(x_{0},0) = \rho(t_1)$, $\tilde{\rho}(s)$ indexed by
    the backward time $s$ evolves under the time-reversed driving
    protocol $\tilde{\lambda}_{s}$.}
 \label{fig:schematic}
\end{figure}

A fundamental relationship in nonequilibrium physics is the Jarzynski
equality \cite{Jarzynski1997April,Jarzynski1997November,Sevick2008May}
which relates the free energy difference between two states of a
system to the work required to force the system from one state to the
other. \added{The work done during a nonequilibrium process is
  described by the time-integral over $\dot{\lambda}_t
  \partial_{\lambda}H$, where $H = H(\lambda_t)$ is a
  $\lambda$-dependent Hamiltonian and $\lambda_t$ is a time-dependent
  control parameter as depicted in Fig.~\ref{fig:schematic}(a).}  The
Jarzynski equality states that starting the system from an equilibrium
distribution, the expectation of the exponential of the negative work
performed on the system is equal to the exponential of the negative
free energy difference between the two states:
\begin{equation}
    \mathbb{E}\big[e^{-\beta W} \big] = e^{-\beta \Delta F}.
    \label{eq:jarzynski}
\end{equation}
Here, $W(t)=\int_0^t \partial_{\lambda}H(z(t),\lambda
(t))\dot{\lambda_{t}}\mathrm{d} t$ is the work performed, $\Delta F$
is the free energy difference, and $\beta = 1/(k_{\rm B} T)$ is the
inverse temperature.

Jarzynski also extended the equality to stochastic trajectories.
His original result was then found to be a consequence of a more general
fluctuation theorem proposed by Crooks \cite{Crooks1999September}. In
this work, Crooks considers the Markovian dynamics of a system that can
be influenced through a time-dependent control parameter $\lambda_t$ and
that satisfies the microreversibility condition
\begin{equation}
    \frac{\mathcal{P}[x_{+t} \mid \lambda_{+t}]}
    {\mathcal{P}[\bar{x}_{-t} \mid \bar{\lambda}_{-t}]}
    =\exp\big(-\beta Q[x_{+t}, \lambda_{+t}]\big).
    \label{eq:crooks-reversibility}
\end{equation}
Here, $\mathcal{P}[x(+t) \mid \lambda_{+t}]$ is the probability density
of the forward trajectory $x_{+t}$, given the control parameter
$\lambda_{+t}$, and $\mathcal{P}[\bar{x}_{-t} \mid \bar{\lambda}_{-t}]$ is
the probability density of the time-reversed trajectory $\bar{x}_{-t}$,
given the time-reversed control parameter $\bar{\lambda}_{-t}$. Starting
from Eq.~\eqref{eq:crooks-reversibility}, Crooks found
\begin{equation}
    \frac{\mathcal{P}_F(W)}{\mathcal{P}_R(-W)} = e^{\beta(W - \Delta F)},
    \label{eq:crooks-fluctuation}
\end{equation}
where $\mathcal{P}_F(W)$ and $\mathcal{P}_R(-W)$ are the probabilities
of observing work $W$ in the forward and reverse processes,
respectively. This result generalizes the Jarzynski equality to a
probability density over work. As with the Jarzynski equality,
derivation of the Crooks fluctuation theorem seems to require an
ensemble of states sampled from equilibrium at the start of the
process.  \added{However, from Jarzynski's work, it was not entirely
  clear how the usual concepts of heat, entropy, and free energy in
  thermodynamics can be defined in a general stochastic system. In
  particular, for what kind of system can the microreversibility
  condition in Eq.~\eqref{eq:crooks-reversibility} be satisfied?}

By introducing the concept of \textit{stochastic energetics}, Sekimoto
\cite{Sekimoto199801} developed a framework \added{connecting
  thermodynamics to overdamped Langevin dynamics with diffusion
  subject to the fluctuation-dissipation relation.} Specifically, the
heat can be computed according to the first law of thermodynamics as
$Q = \Delta U - W$, where $\Delta U$ is the change in internal energy.

\added{Later, Seifert \cite{Seifert2005July} showed that
  microreversibility holds in overdamped Langevin dynamics obeying the
  fluctuation-dissipation relation. Further extending these results,
  he} explicitly introduced a trajectory-dependent,
information-theoretic formulation of entropy production into
nonequilibrium thermodynamics, which was also used implicitly in
\cite{Crooks1999September} and discussed in
\cite{Qian2001December}. The trajectory-dependent entropy was defined
as
\begin{equation}
    S(x_{t},t) \equiv  -k_{\rm B} \ln \rho(x_{t},t),
    \label{eq:entropy}
\end{equation} 
where $\rho(x_{t},t)$ is the probability density \added{in state
  coordinate $x$ evaluated at the value $x_{t}$ of the stochastic
  trajectory $\{x_t: t \geq 0\}$ at time $t$.} In
Eq.~\eqref{eq:entropy}, the probability density is implicitly defined
relative to a baseline density $\rho(x_{0},0)$.  Seifert then
considered overdamped Langevin dynamics under the influence of a
Hamiltonian (potential) $V(x,\lambda)$ and an external force
$f(x,\lambda)$, both subject to a time-dependent control parameter
$\lambda_t$. In the stochastic differential equation (SDE)
formulation, the associated dynamics obey
\begin{equation}
  \mathrm{d}x_{t} = \mu\big[-\partial_x V(x,\lambda_{t})
    +f(x_{t}, \lambda_{t})
 \big] \mathrm{d}t + \sqrt{2D}\, \mathrm{d}B_{t},
\end{equation}
where $\mu$ is the mobility, $D$ is the diffusion coefficient, and
$B_{t}$ is a Wiener process. Using the path integral formulation of
overdamped Langevin dynamics \cite{Kurchan1998April}, Seifert
validated Eq.~\eqref{eq:crooks-reversibility} and
Eq.~\eqref{eq:crooks-fluctuation}.  Importantly, he noted that the
Eq.~\eqref{eq:crooks-reversibility} and
Eq.~\eqref{eq:crooks-fluctuation} can hold without the
  requirement that both the initial and final states are in
equilibrium.  In fact, the Crooks fluctuation theorem holds for finite
time and any initial distribution. Eq.~\eqref{eq:jarzynski} was then
generalized to the following form:
\begin{equation}
    \mathbb{E}\big[e^{-\beta (W_t -\Delta F_t)} \big] = 1.
    \label{eq:seifert}
\end{equation}
Note that we have adapted the notation in \cite{Seifert2005July} to be
consistent with that in the rest of this paper. In the original paper,
the heat exchange between the system and the environment is understood 
in terms of entropy change in the environment. Then, considering the
total entropy change in the system and the environment, we have 
$\Delta S_{\rm tot} =  (W- \Delta F)/T$.

Subsequently, Sagawa and Ueda \cite{Sagawa2010March} introduced the
concept of \textit{feedback control} and \textit{measurement} into
nonequilibrium thermodynamics, generalizing the Jarzynski equality to
account for systems where a feedback controller influences the
dynamics.  They further generalized the Jarzynski equality to 
\begin{equation}
    \mathbb{E}\big[e^{-\beta(W_t - \Delta F_t) - I_t} \big] = 1,
\end{equation}
where $I_t$ is the (trajectorywise version of) mutual information
between the actual system state $x$ and the measurement outcome $y$,
defined to be $I_t=I\big(x_t,y_t,t\big) = - \ln
\left[{\rho\big(x_t,y_t,t\big)}/{[\rho\big(x_t,t\big)
      \rho\big(y_t,t\big)]}\right]$. This is the first result that
explicitly connects the Jarzynski equality to information theory.


More recently, martingale properties of entropy production in
stochastic systems have been explored by Neri \cite{Neri2020January} and
Manzano et al. \cite{Manzano2021February}, who extended the Jarzynski
equality from a fixed time $t$ to a stopping time $\tau \leq t$. Through
the use of path probability densities from a path integral formulation,
they derived the following martingale that is associated with entropy
production of an overdamped Langevin system:
\begin{equation}
    \mathbb{E}\big[ e^{-\beta(W_{\tau} - \Delta F_{\tau}) - \delta_{\tau}} \big] = 1,
    \label{eq:manzano}
\end{equation}
where the stochastic distinguishability $\delta_{\tau}$ is defined as
\begin{equation}
  \delta_{\tau} = \ln \left[
    \frac{\rho(x_{\tau},\tau)}{\tilde\rho{(x_{\tau}, t-\tau)}}\right].    
    \label{eq:delta}
\end{equation} 
Here, $\rho(x_{\tau}, \tau)$ is the probability density of the system
at $x$, evaluated at $x_{\tau}$ at time $\tau$, and $\tilde{\rho}$ is
the time-reversed probability density under a time-reversed driving
protocol as shown in Fig.~\ref{fig:schematic}(b,c).

Nearly all previous results have been developed using a path integral
formulation \cite{Kurchan1998April} for overdamped Langevin dynamics
or quantum systems \cite{Manzano2021February,Santiago2024Apr}.  While
the path integral formulation is a convenient tool in many areas of
physics, gaps in its mathematical rigor may preclude certain desirable
directions of analysis. For example, the path integral integrates over
the space of continuously differentiable functions but solutions to
stochastic differential equations (SDEs) are nowhere
differentiable. There have been several efforts to formulate the path
integral in a mathematically rigorous way
\cite{albeverio1997wiener,mazzucchi2009mathematical,
  albeverio2022mathematical}; however, these approaches were primarily
focused on quantum path integrals and typically assigned a different
interpretation of the probability density in the path integral.  A
mathematically satisfying formulation of the path integral for
classical stochastic systems can arise through Girsanov's theorem, and
was treated previously in
\cite{dutta2023fluctuationtheoremspecialcase}.  Even though Girsanov's
theorem is a powerful tool in the theory of stochastic calculus,
dealing with the path integral with both forward and backward paths is
challenging in terms of \added{precise interpretation of the
  probability density in the forward and backward paths.  Moreover,
  derivations using path integrals rely on microreversibility,
  precluding treatment of far-from-equilibrium systems where
  microreversibility does not hold.}

On the other hand, solving the corresponding Fokker-Planck equation to
find the trajectorywise entropy $S(x_t,t)$ at sufficient numerical
precision requires significant computational resources.  This is
especially challenging for high-dimensional systems or systems with
complex potentials. While biological systems can often be described as
Maxwell's demons that convert information into work
\cite{Brown2019August,Leighton2024November}, the aforementioned
computational demands limits application of the generalized work
theorem to biological systems.  Thus, experimental verification of the
Jarzynski equality has been restricted to relatively simple artificial
systems \cite{Toyabe2010November,An2014December,Paneru2020February}.

In this paper, we side-step path integration by providing an
alternative mathematical proof of the martingale property of entropy
production.  While our method mirrors that described in a recent
treatise on martingale methods for physicists \cite{Roldn2023April},
we further show that this proof reveals a generalization to the work
theorem, extending it to a family of equations that hold for the same
stochastic process but using different choices for the backward
process.
%
%
Our proof also strengthens the equality in Eq.~\eqref{eq:manzano} by
explicitly evaluating the conditional expectation of the same
exponential given the initial value $x_0$ for \textit{any} initial
distribution $\rho_0$. Thus, it is not necessary to average over all
trajectories sampled from the initial distribution $\rho(x_0,0)$.
%
%
Moreover, our new ``forward-backward-decoupled'' work theorem can be
developed using underdamped dynamics described by position $x$ and
velocity $v$. Our generalized work theorem can be applied to
high-dimensional out-of-equilibrium systems or systems with complex
potentials with lower computational costs.

\newpage

\section{Analysis and Results}
We formally derive and and describe a number of results below.

\subsection{Mathematical Approach}
We provide the mathematical intuition behind our approach. Consider a
stochastic process $x$ that evolves according to a stochastic
differential equation (SDE) of the form $\mathrm{d}x_t =
b_t\mathrm{d}t + \sigma_t\mathrm{d}B$. If $b_t\equiv 0$, the process
$x_t$ is purely driven by diffusion and has no ``bias'' to drift in
any direction.  Once some regularity conditions are satisfied, the
mean of the process $x_t$ is the same as that of the initial
condition, \textit{i.e.}, $\mathbb{E}[x_t] = \mathbb{E}[x_0]$.

We formalize the intuition developed above by introducing the
concept of an exponential martingale. Consider a predictable process
$\theta_t$ adapted to the filtration of a standard Wiener process $B_t$.
The \textbf{exponential martingale} $(M_t)_{t \geq 0}$ associated with
$\theta_t$ is defined by:
\begin{equation}
  M_t \equiv \exp\Big( \medint\int_0^t \!\theta_s
  \, \textrm{d}B_s - \mfrac{1}{2} \medint\int_0^t \!\theta_s^2 \, 
    \textrm{d}s \Big),
    \label{eq:exp_mart}
\end{equation}
where $\int_0^t \theta_s \,\textrm{d}B_s$ denotes the stochastic
integral with respect to the Wiener process. The term $\tfrac{1}{2}
\int_0^t \theta_s^2 \, \textrm{d}s$ is the compensating drift term,
ensuring that $M_t$ has zero mean drift.

To guarantee that $\big[M_t\big]_{t \geq 0}$ is indeed a true
martingale, a sufficient condition is provided by the well-known
\textbf{Novikov Condition}, which states that if
\begin{equation}
  \mathbb{E}\bigg[\exp\Big(\mfrac{1}{2}
    \medint\int_0^T \!\theta_s^2 \, \textrm{d}s
    \Big)\bigg] < \infty \quad \forall\, T > 0,
    \label{eq:novikov}
\end{equation}
holds then the process $M_t$ defined in \eqref{eq:exp_mart} is a true
martingale. Throughout this paper, we will use It\^o calculus
rules for stochastic integrals. Interested readers can refer to
\cite{karatzas2014brownian} for a comprehensive and pedagogical
introduction to stochastic calculus.

\subsection{Overdamped Dynamics}

Consider overdamped Langevin dynamics defined by
\begin{align}
    \gamma \mathrm{d} x_t &= -\nabla H(x_t,t) \, \mathrm{d} t +
    f(x_t,t) \,\mathrm{d} t + \sqrt{\mfrac{2 \gamma}{\beta}}\,
    \mathrm{d} B_t.
    \label{eq:overdamped}
\end{align}
Here, $x$ is the state of the system, $H(x,t)$ is the Hamiltonian,
$f(x,t)$ is the external force, and $\gamma$ is the friction
coefficient.  The time dependence of $H(x,t)$ and $f(x,t)$ can include
that of the control parameter $\lambda_t$ in the original formulation.

The corresponding Fokker-Planck equation is given by
\begin{align}
    \partial_t \rho(x,t) = \mfrac{1}{\gamma} 
    \nabla \cdot \left[ \big(\nabla H(x,t) - f(x,t)\big) \rho(x,t) \right]  
    + \mfrac{1}{\beta \gamma} \Delta \rho(x,t),
    \label{eq:fokker}
\end{align}
where $\rho(x,t)$ is the probability density of the system at time
$t$, and $\nabla$ and $\Delta$ are the gradient and Laplacian
operators, respectively, with respect to the state variable $x$.

Along a given trajectory $x_{s\leq t}$, the work performed on the
system up to time $t$ is given by \cite{Qian2001December}
\begin{equation}
    \begin{aligned}
        W_t & = \int_0^t f(x_s,s) \circ \mathrm{d} x_s 
        + \int_0^t \partial_{t} H(x_s,s) \,\mathrm{d}s,\\
        \: & = \int_0^t f(x_s,s) \,\mathrm{d}x_s
        +  \mfrac{1}{\gamma \beta}\int_0^t \nabla \cdot f(x_s,s) \,\mathrm{d}s 
        + \int_0^t  \partial_{t}H(x_s,s) \,\mathrm{d}s,
    \end{aligned}
    \label{eq:work}
\end{equation}
where $\circ$ denotes the Stratonovich integral. The entropy $S(x_t)$
specific to the trajectory $x_t$ is given by Eq.~\eqref{eq:entropy}.

\paragraph{Our backward process.} While the identity
  in Eq.~\eqref{eq:seifert} suggests that $\exp \big[-\beta (W -
    \Delta F)\big]$ is a martingale, it is actually not if $F = H - T
  S$. We will have to define a backward process and use the
  trajectorywise entropy of the backward process to construct a
  martingale.

Consider a time-reversed driving protocol described by
$\tilde{\lambda}_{s}$ depicted in Fig.~\ref{fig:schematic}(b). In our
analysis we will implicitly define forces, Hamiltonians, and densities
under the reversed protocol by reversing the time direction of the
Fokker-Planck equation by switching the sign of the time-derivative
term and do not need to explicitly invoke
$\tilde{\lambda}_{s}$. Specifically, the Hamiltonian and external
force are replaced by their time-reversed forms,
$\widetilde{H}(x,s)\coloneqq H(x,t_1-s)$ and
$\widetilde{f}(x,t)\coloneqq f(x,t_1-s)$
\cite{Manzano2021February}. Here, $s = t_1 - t$ represents how far
back in time the current time is compared to the terminal time
$t_1$. A new ``time-reversed'' probability density $\tilde{\rho}(x,s)$
with initial condition $\tilde{\rho}(x,0) = \rho(x,t_1)$, depicted in
Fig.~\ref{fig:schematic}(c), evolves according to the ``backward''
Fokker-Planck equation

\begin{equation}
      \partial_t \tilde{\rho} = \mfrac{1}{\gamma} \nabla
      \cdot \left[ \big(\nabla \widetilde{H}(x,t)
        - \widetilde{f}(x,t)\big) \tilde{\rho} \right]
      + \mfrac{1}{\beta \gamma} \Delta \tilde{\rho}.
      \label{FP_tilde_rho}
\end{equation}
When the potential $H(x,t)$ and the force $f(x,t)$ are
time-asymmetric, i.e., $H(x,t) \neq \widetilde{H}(x,t)$ and $f(x,t)
\neq \widetilde{f}(x,t)$, the time-reversed process for
$\tilde{\rho}(x,t)$ is different from the original process for
$\rho(x,t)$ in the sense that $\tilde{\rho}(x,t) \neq \rho (x, t_1-t)$
for $t \in (0,t_1)$. Note that the Fokker-Planck equation is also
known as the Kolmogorov forward equation, but here, the backward
Fokker-Planck equation differs from the Kolmogorov backward equation
which is simply the adjoint of the Kolmogorov forward equation
\cite{ANDERSON1982313}.
%
%
%
%

We define our backward process by its probability density $\psi(x,t)$
which obeys a specific type of backward Fokker-Planck equation
\begin{equation}
  -\partial_t \psi{(x,t)} = \mfrac{1}{\gamma}
  \nabla \cdot \Big[ 
        \big(\nabla H(x,t) - f(x,t)\big) \psi(x,t) \Big]  
    + \mfrac{1}{\beta \gamma} \Delta \psi(x,t).
    \label{eq:backward}
\end{equation}
Eq.~\eqref{eq:backward} differs from the Fokker-Planck
Eq.~\eqref{eq:fokker} by an extra minus sign in front of the time
derivative. It is straightforward to verify that $\psi(x,t) =
\tilde{\rho}(x,t_1 -t)$ and that $\psi(x,t)$ is associated with the
time-reversed probability $\tilde{\rho}$ used in
\cite{Manzano2021February}.


In the setting of Crooks' fluctuation theorem
\cite{Crooks1999September}, a specific terminal time $t_1$ was
\textit{a priori} chosen, and $\psi(x,t) \coloneqq \tilde{\rho}(x,t_1
- t)$ solves Eq.~\eqref{eq:backward} with initial condition
$\psi(x,t_1) = \rho(x,t_1)$. The specific time-reversed probability
density $\tilde{\rho}(x,s)$ is used in Eq.~\eqref{eq:delta} to define
the stochastic distinguishability $\delta$. $\psi(x,t_1 -s)
=\tilde{\rho}(x, s)$ maps the time-reversed probability density at the
backward time $s$ to the time-reversed probability density at the
forward time $t = t_1 -s$, as illustrated in
Figs.~\ref{fig:schematic}(c).

\begin{figure}[htb]
  \centering
  \includegraphics[width = 6.4in]{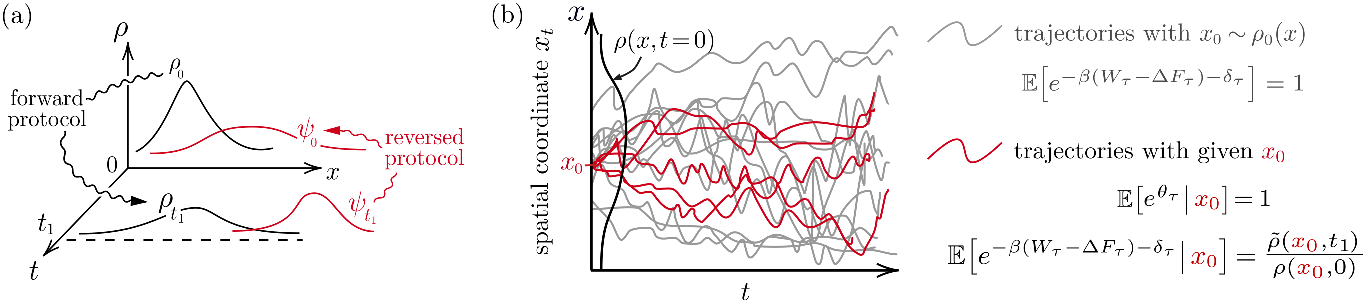}
  \caption{Decoupling of the forward and backward processes.  (a) In
    our derivations, we employ the probability density $\psi(x,t)$ of
    a backward process indexed by forward time $t$.  The initial
    condition is arbitrary so that in general $\psi(x,t_1) \neq
    \rho(x,t_1)$. (b) Trajectories of the forward processes $x_t$
    sampled from the initial distribution $\rho(x_{0},0)$ are shown in
    grey, while trajectories with a specific initial value $x_0$ are
    shown in red. The original work theorem uses averages over the
    grey trajectories, while our generalized work theorem considers
    averages over the red trajectories.  See Eqs.~\eqref{eq:main} and
    \eqref{eq:manzano2}.}
  \label{fig:schematic2}
\end{figure}

To derive the generalized work theorem, $\psi(x,t)$ need only satisfy
Eq.~\eqref{eq:backward} regardless of initial condition.  Our
generalization of the work theorem is based on the observation that
the backward process $\psi(x,t)$ can be defined with an arbitrary
initial condition, not necessarily $\rho(x,t_1)$, as shown in
Fig.~\ref{fig:schematic2}(a). Broadly speaking, there is no need to
find a specific terminal time $t_1$ in our approach.

Analogous to the entropy $S(x_t)$ defined in Eq.~\eqref{eq:entropy},
we define the ``entropy'' of the backward process along the trajectory
$x_t$ as
\begin{equation}
        \Sigma(x_t,t) = -k_{\rm B}\ln \psi(x_t,t).
  \label{SIGMA}
\end{equation}
As with the trajectorywise entropy, $\Sigma(x_t,t)$ implicitly
requires a reference probability density $\psi(x_0,0)$.  Throughout
this work, we use $A_t$ to denote the trajectory-specific value of
$A(x,t)$ evaluated at $x=x_t$, e.g., $\Sigma_t \equiv \Sigma(x_t,t)$.

\paragraph{Martingale and generalized work theorem.}
We now consider the quantity defined by
\begin{equation}
 \theta_t = - \beta W_t + \beta \Delta H(x_t,t)-\beta T \Delta \Sigma(x_t,t).    
  \label{THETA_SIGMA}
\end{equation}
This process represents the nondimensionalized ``entropy production''
along the trajectory $x_t$ when the entropy $S$ is replaced by that of
the backward process $\Sigma$.  Upon using It\^{o}'s formula to expand
terms, the time derivative of $\theta_t$ is given by
\begin{equation}
\begin{aligned}
    \mathrm{d} \theta_t & = - \beta f \cdot \mathrm{d}x_t 
    - \mfrac{1}{\gamma} \nabla \cdot f\, \mathrm{d}t 
    - \beta \partial_{t} H\, \mathrm{d}t +
    \beta \nabla H \cdot \,\mathrm{d}x_t 
    + \mfrac{\beta}{2}\Delta H \,\big[\mathrm{d}x_t\big]^2 
    + \beta \partial_{t} H\, \mathrm{d}t 
    + \frac{\mathrm{d}\psi}{\psi}
    - \frac{[\mathrm{d} \psi]^2}{2 \psi^2}\\
    \:  & = \beta(\nabla H -f) \cdot \mathrm{d}x_t +
    \mfrac{1}{\gamma} (\Delta H - \nabla \cdot f)\,\mathrm{d}t 
    + \frac{\mathrm{d}\psi}{\psi}
    - \frac{[\mathrm{d} \psi]^2}{2 \psi^2}.
    \label{eq:mid0}
\end{aligned}
\end{equation}

Our goal is to express $\mathrm{d} \theta_t$ in terms of $\mathrm{d}
B_t$ and $\mathrm{d} t$. To achieve this, we need to evaluate
$\mathrm{d}\psi$ as $\mathrm{d}\psi(x_t,t)$.
Using the chain rule and Eq.~\eqref{eq:backward}, we find
\begin{equation}
    \begin{aligned}
      \mathrm{d} \psi &= \partial_t \psi \,\mathrm{d}t
      + \nabla \psi \cdot \mathrm{d}x_t
      + \mfrac{1}{2} \Delta \psi \, [\mathrm{d}x_t]^2\\
      \:  &= -\mfrac{1}{\gamma} \nabla \cdot
      \big[(\nabla H - f) \psi \big] \mathrm{d}t 
        -\mfrac{1}{\beta \gamma} \Delta \psi \,\mathrm{d}t 
        + \nabla \psi \cdot \mathrm{d}x_t
     + \mfrac{1}{2} \Delta \psi [\mathrm{d}x_t]^2\\
     \:  &= - \mfrac{\psi}{\gamma}
     (\Delta H - \nabla \cdot f) \, \mathrm{d} t
     - \mfrac{1}{\gamma}\big[(\nabla H - f) \cdot \nabla \psi\big]
     \mathrm{d}t  + \nabla \psi \cdot \mathrm{d}x_t 
    \end{aligned}        
\end{equation}
Consequently, $\tfrac{\mathrm{d}\psi}{\psi}$ and $\tfrac{1}{2
  \psi^2}[\mathrm{d} \psi]^2$ are given by
\begin{equation}
    \begin{aligned}
      \mfrac{1}{\psi}\, \mathrm{d} \psi & =
      - \mfrac{1}{\gamma} (\Delta H - \nabla \cdot f) \,\mathrm{d} t
      - \mfrac{1}{\gamma} \big[(\nabla H - f)
        \cdot \nabla \ln \psi \big] \mathrm{d}t + 
        \nabla \ln \psi \cdot \mathrm{d}x_t \\
        \mfrac{1}{2 \psi^2} [\mathrm{d}  \psi]^2 & =
        \mfrac{1}{\gamma \beta}\mfrac{\| \nabla \psi\|^2}{\psi^{2}}\mathrm{d}t.
    \end{aligned}
    \label{eq:mid1}
\end{equation}
Substituting Eqs.~\eqref{eq:mid1} into Eq.~\eqref{eq:mid0}, we find
\begin{equation}
    \begin{aligned}
      \mathrm{d} \theta_t & = - \mfrac{1}{\gamma}
      (\nabla \ln \psi) \cdot (\nabla H -f) \,\mathrm{d} t
      - \mfrac{1}{\gamma \beta} \big\|\nabla \ln \psi\big\|^2 \,\mathrm{d}t
      +        \big( \beta \nabla H - \beta f + \nabla \ln \psi \big)
      \cdot \mathrm{d}x_t\\
      \: & =- \mfrac{1}{\gamma \beta} \big\| \beta (\nabla H -f)
      + \nabla \ln \psi \big\|^2 \,\mathrm{d}t +
      \sqrt{\mfrac{2}{\gamma \beta}}\,\Big[\beta (\nabla H -f)
        + \nabla \ln \psi \Big] \cdot \mathrm{d}B_t.
    \end{aligned}
    \label{eq:theta_overdamped_final}
\end{equation}
The exponential of $\theta_t$ can now be expressed as 
\begin{equation}
    \begin{aligned}
      \mathrm{d} e^{\theta_t} & = e^{\theta_t}
      \left[\mathrm{d} \theta_t
        + \mfrac{1}{2} (\mathrm{d}\theta_t)^2 \right]\\
      \: & = e^{\theta_t} \sqrt{\mfrac{2}{\gamma \beta}}\,
      \Big[\beta (\nabla H -f)
        + \nabla \ln \psi \Big] \cdot \mathrm{d}B_t.       
    \end{aligned}
    \label{dtheta1}
\end{equation}
Thus, as long as the Novikov condition holds, the process
$e^{\theta_t}$ is a martingale with initial value one.

By the optional stopping theorem for martingales, we have
\begin{equation}
  \mathbb{E} \left[e^{\theta_{\tau}} \mid x_0\right]
  = e^{\theta_{0}} = 1, ~\text{a.s.},~ \forall
  ~\text{bounded stopping time}~\tau.
\label{eq:main}
\end{equation}
Eq.~\eqref{eq:main} is the main result of this paper; note that
Eq.~\eqref{eq:main} is stronger than
\added{$\mathbb{E}\left[e^{\theta_\tau}\right] = 1$} since
Eq.~\eqref{eq:main} is an average over trajectories starting from an
arbitrary initial value $x_0$, independent of the initial distribution
$\rho_0(x)$, while the latter is an average over all trajectories
starting from the initial distribution $\rho_0(x)$, shown by the red
and grey trajectories in Fig.~\ref{fig:schematic2}(b), respectively.

\paragraph{Manzano's result.}
We now show that Eq.~\eqref{eq:main} is a generalization of
Eq.~\eqref{eq:manzano}. The key observation is that in the definition
of the backward process $\psi(x,t)$, we have the freedom to choose the
initial condition. For different choices of the initial condition, we
will arrive at different forms of the martingale.  \added{Specifically, in
the setting of \cite{Manzano2021February}, the backward process is
chosen by fixing a final time $t_1$, and choosing the ``initial
condition'' of the backward process to be
\begin{equation}
    \psi(x,t_1) = \rho(x,t_1),\,\,  \forall x.
\end{equation}
Then, $\psi(x,t)$ evolves backward in time according to
Eq.~\eqref{eq:backward}.}



Consider a stopping time $\tau$ such that $0 \leq \tau \leq t_1$
almost surely. Note that $-\beta T \Delta \Sigma (\tau) = \ln
\psi(x_\tau,\tau) - \ln \psi(x_0,0) = \ln \tilde{\rho}(x_\tau,t_1 -
\tau) - \ln \tilde{\rho}(x_0,t_1)$, while $-\beta T \Delta S = \ln
\rho(x_\tau,\tau) - \ln \rho(x_0,0)$. Recalling the definition of
$\delta_{\tau}$ in Eq.~\eqref{eq:delta}, we find
\begin{equation}
    \mathbb{E}\Big[e^{- \beta \big[W_{\tau} - \Delta F_{\tau}\big]- \delta_\tau  } \Big] = 
    \mathbb{E} \Big[ e^{\theta_{\tau}}
      \frac{\tilde \rho(x_0,t_1)}{\rho(x_0,0)}\Big]
    = \int \tilde{\rho}(x_0,t_1)  \mathbb{E} 
    \left[ e^{\theta_t} \mid x_0 \right]  \mathrm{d} x_0 = 1,
\end{equation} 
which is Eq.~\eqref{eq:manzano}. Here, the second equality follows
from conditioning on $x_0$ and the last equality follows from
Eq.~\eqref{eq:main}.  Additionally, by conditioning on $x_0$, we have
\begin{equation}
  \mathbb{E}\Big[e^{- \beta \big[W_{\tau} - \Delta F_{\tau}\big]- \delta_\tau}
    \Big|~ x_0\Big] = 
    \mathbb{E} \bigg[ e^{\theta_t} \frac{\tilde \rho(x_0,t_1)}
    {\rho(x_0,0)}\bigg| ~x_0 \bigg]
    = \frac{\tilde \rho(x_0,t_1)}{\rho(x_0,0)}, \quad \forall x_0.
    \label{eq:manzano2}
\end{equation} 
Eq.~\eqref{eq:manzano2} is stronger than Eq.~\eqref{eq:manzano} since
Eq.~\eqref{eq:manzano} only holds for all trajectories sampled
according to the initial distribution $\rho_0(x)$, while our result
holds for trajectories starting from an arbitrary initial value $x_0$,
independent of the initial distribution $\rho_0(x)$, as shown in
Fig.~\ref{fig:schematic2}(b).

This result is particularly helpful when one wants to consider
the work theorem for a system where the initial distribution cannot
be written as a density function over the state space, such as the
Dirac delta distribution. In such cases, the trajectorywise entropy
at the initial time $S(0) = - k_{\rm B} \ln \rho(x_0,0)$ is not
well-defined.  Our result provides a way to bypass this issue by
decoupling the initial sample $x_0$ from its initial distribution
$\rho_0$. Energy changes and external work can be evaluated by the
conditional work theorem Eq.~\eqref{eq:manzano2} for trajectories
starting at $x_0$ with an arbitrarily chosen well-behaved initial
distribution $\rho_0$.

\paragraph{Stationary Hamiltonian.}
There are other interesting choices for the backward process. For
example, in the case of time-independent potentials and forces, we can
choose $\psi(x,0) = \rho_*(x)$, the stationary distribution of
Eq.~\eqref{eq:fokker}. In this case, $\psi(x,t) \equiv \psi(x,0) =
\rho_*(x)$ and
\begin{equation}
    \mathbb{E} \Big[e^{-\beta \big(W_{\tau} - \Delta H_{\tau} + T \Delta S_*(\tau) \big)}\Big] = 1, 
    \,\, \forall ~\text{bounded stopping time}~\tau,
    \label{eq:stationary}
\end{equation}
where $S_*(t) \coloneqq -k_{\rm B}\ln \rho_*(x_t)$ and the initial
distribution $\rho_0(x)$ of $x_0$ can be different from $\rho_*(x)$.

The stationary case is of interest when $f$ is a dissipative
force. Such a system can be used to model nonequilibrium stochastic
chemical reactions commonly found in biological systems, including
kinetic proofreading \cite{Hopfield1974oct,li2024reliable} and
chemotaxis \cite{hathcock2023nonequilibrium}. While evaluating
Eq.~\eqref{eq:manzano} requires solution to the $d+1$-dimensional
time-dependent PDE for $\rho(x,t)$, computing
Eq.~\eqref{eq:stationary} requires only the solution to the
$d$-dimensional time-independent PDE for $\rho_*(x)$, leading to an
easier computational evaluation. A similar identity was derived in
\cite{Roldn2023April} assuming that the initial distribution
$\rho_0(x)$ is the stationary (backward) distribution $\rho_*(x)$. Our
result relaxes this assumption and shows that the identity holds for
any initial distribution $\rho_0(x)$.


\paragraph{Fluctuation-dissipation relation.}
In our formulation of the Langevin dynamics in
Eq.~\eqref{eq:overdamped}, we related the diffusion coefficient
$D=1/(\beta \gamma)$ to the friction coefficient $\gamma$ as
required by the fluctuation-dissipation theorem. Specifically, a
general overdamped process is governed by
\begin{equation}
  \mathrm{d}x_t = \mfrac{1}{\gamma}
  \big[-\nabla H(x_t,t) + f(x_t,t)\big]\,\mathrm{d}t 
    + \sqrt{2D}\,\mathrm{d}B_t.
  \label{eq:overdamped_general}
\end{equation}
Eq.~\eqref{eq:overdamped} is obtained from
Eq.~\eqref{eq:overdamped_general} by setting $D = 1/(\beta \gamma)$.

Starting from Eq.~\eqref{eq:overdamped_general}, using the backward
process $\psi(x,t)$ associated with the time-reversed Fokker-Planck
Eq.~\eqref{eq:overdamped_general}
\begin{equation}
  -\partial_t \psi(x,t) 
  = \mfrac{1}{\gamma} \nabla \cdot \Big[ \big(\nabla H(x,t)
    - f(x,t)\big) \psi(x,t) \Big] 
  + D \Delta \psi(x,t),
\end{equation}
and defining $\Sigma_t, \theta_t$ through Eqs.~\eqref{SIGMA} and
\eqref{THETA_SIGMA}, we find (see Appendix~\ref{app:general_derivation}
for a detailed derivation)
\begin{equation}
  \begin{aligned}
    \mathrm{d}e^{\theta_t} = e^{\theta_t} \bigg[\Big(D - \mfrac{1}{\beta \gamma} \Big)
      \Big( \beta^2 & \big\|\nabla H - f\big\|^2
      + 2\beta  \big(\nabla H - f\big) \cdot \nabla \ln \psi +
      \beta \big(\Delta H - \nabla \cdot f\big) \Big) \mathrm{d}t \\
      \: &  + \sqrt{2D} \Big( \beta \big(\nabla H - f\big)
      + \nabla \ln \psi \Big) \cdot \mathrm{d}B_t\bigg].
  \end{aligned}
\end{equation}
When the assumption $D=1/(\beta \gamma)$ is violated,
$\mathrm{d} e^{\theta_t}$ carries a non-zero drift and is no longer a
martingale, unless $f = \nabla H$ cancelling out the energy gradient
everywhere.


\subsection{Generalization to underdamped Langevin dynamics}
Our method of introducing an backward process is general and can be
applied to \textit{underdamped} Langevin dynamics that obey
\begin{equation}
    \begin{aligned}
        \displaystyle
        {\mathrm{d} x_{t}} & = v_{t} \,{\mathrm{d} t}\\
        \displaystyle
        m\mathrm{d} v_{t} & = \big[-\gamma v_{t} +
        (-\nabla_x U+f)\big] \mathrm{d} t
        + \sqrt{\mfrac{2 \gamma}{\beta }} \, \mathrm{d} B_t,
    \end{aligned}
    \label{eq:underdamped}
\end{equation}
where $m$ is an effective mass and $\gamma$ represents a velocity
decay rate, or the decay rate of the velocity-velocity correlation
function under fluctuation-dissipation conditions.

The main differences in analyses between the underdamped and overdamped
cases are the definition of the Hamiltonian
\begin{equation}
    H(x,v,t) = \mfrac{1}{2} m \| v\|^2 + U(x,t),
\end{equation}
and the evaluation of the work performed
\begin{equation}
      W_t = \int_0^t f(x_s,s) \,\mathrm{d}x_s
      + \int_0^t \partial_{t}U(x_s,s) \,\mathrm{d}s.
\end{equation}
Here, because $x_s$ is now a continuously differentiable function of
$s$, the Stratonovich integral with respect to $x_s$ is equivalent to
the normal Lebesgue-Stieltjes integral. The evolution of the
probability density $\rho(x,v,t)$ is then given by

\begin{equation}
    \partial_t \rho(x,v,t) = -v \cdot \nabla_x \rho - \mfrac{1}{m} \nabla_v \cdot \Big[
        \big(- \gamma v - \nabla_x U +f\big) \rho
    \Big] + \mfrac{\gamma }{\beta m^2} \Delta_v \rho.
    \label{kramers}
\end{equation}
where we use $\nabla_x$ and $\nabla_v$ to denote the gradient
with respect to $x$ and $v$, respectively, and $\Delta_v$ to represent
the Laplacian with respect to $v$. Correspondingly, we define a
backward process $\psi(x,v,t)$ by
\begin{equation}
      -\partial_t \psi(x,v,t) = v \cdot \nabla_x \psi
      - \mfrac{1}{m} \nabla_v \cdot \Big[
            \big(- \gamma v + \nabla_x U -f\big) \psi
        \Big] + \mfrac{\gamma}{\beta m^2} \Delta_v \psi.
      \label{psi_underdamped}
\end{equation}
Eq.~\eqref{psi_underdamped} is obtained by applying the
transformations $v \to -v$, $\nabla_v \to - \nabla_v$ and $\partial_t
\to - \partial_t$ to the forward Kramers equation~\eqref{kramers}.
$\psi(x,v,t)$ represents a probability density in the sense that
$\tilde{\rho}(x,v,t)\coloneqq \psi(-x,v,-t)$ is the probability
density of the Langevin dynamics under an inverted and
  time-reversed potential and force $\widetilde{U}(x,t)=U(-x,-t)$ and
  $\tilde{f}(x,t)=-f(-x,-t)$.  Specifically, we do not impose
constraints on the initial condition of $\psi(x,v,0)$.

The backward process $\psi(x,v,t)$ or $\tilde{\rho}(-x,v,-t)$ is a
generalization of the previous ``time-reversed'' processes discussed
in \cite{Manzano2021February} for underdamped Langevin dynamics. To
make the connection clearer, we can choose a specific final time $t_1$
and shift the time argument to $t_1 - t$ in the backward process,
\textit{i.e.}, $\tilde{\rho}(x,v,t) = \psi(-x,v,t_1 - t)$. If we let
$\tilde{\rho}(x,v,0) = \psi(-x,v,t_1)$, $\tilde{\rho}$ evolves forward
in time under the space- and time-reversed potential $\widetilde{U}$
and force $\tilde{f}$ associated with the time-reversed driving
protocol, analogous to that shown in Fig.~\ref{fig:schematic}(b).

\added{Ignoring their dependence on $v_{t}$, the evolution of
  distributions for forward and backward processes is analogous to
  that of the overdamped case depicted in
  Fig.~\ref{fig:schematic2}(a).} Note that we also impose space
inversion which is different from the standard backward process in
which only the time and velocity are reversed. In the latter case of
velocity inversion, the term $\gamma (\nabla _v\cdot v)/m$ carries the
opposite sign, rendering the exponential of the ``entropy'' production
(defined below) no longer a martingale.  Moreover, in our
generalization of the work theorem, we do not impose constraints on
the initial or final condition of $\psi(x,v,t)$.  An arbitrary choice
of the initial condition enables decoupling of the forward and
backward processes.

Extending the quantities in Eqs.~\eqref{SIGMA} and
\eqref{THETA_SIGMA}, we define the entropy production along
trajectories $(x_t,v_t)$ as $\theta_t = - \beta W_t + \beta
H(x_t,v_t,t) + \ln \psi(x_t,v_t,t)$. Chain-rule-expanding
the all terms, we find
\begin{equation}
    \begin{aligned}
        \mathrm{d} \theta_t = & \beta v \cdot (\nabla _x H - f)
        \mathrm{d}t + \beta \nabla _v H \cdot \mathrm{d}v_t +
        \mfrac{\gamma}{m^2} \Delta_v H \mathrm{d}t  + v
        \cdot \nabla_x \ln \psi \,\mathrm{d}t \\
        \: & \qquad + \nabla_v \ln \psi
        \cdot \mathrm{d}v_t + \mfrac{\gamma}{\beta m^2}
        \mfrac{\Delta_v \psi}{\psi} \,\mathrm{d} t
        + \mfrac{\partial_{t}\psi}{\psi}\mathrm{d}t -
        \mfrac{\gamma}{\beta m^2} \big\| \nabla_v \ln \psi \big\|^2
        \mathrm{d} t.
    \end{aligned}
    \label{dtheta0}
\end{equation}
%
If $U$ and $f$ are independent of $v$, substituting $\partial_t \psi$
from Eq.~\eqref{psi_underdamped} into Eq.~\eqref{dtheta0}, we find
\begin{equation}
    \begin{aligned}
        \mathrm{d} \theta_t = &  \beta v \cdot (\nabla_x U - f)  \mathrm{d}t 
        + \beta \nabla_v H \cdot \mathrm{d}v_t
        + \mfrac{n \gamma}{m} \,\mathrm{d}t \\
       \:  & \quad + \nabla_v \ln \psi \cdot \mathrm{d}v_t
        - \mfrac{n \gamma}{m}\, \mathrm{d}t 
        + \mfrac{1}{m} (\nabla _x U -f) \cdot \nabla _v \ln \psi \,\mathrm{d} t
        - \mfrac{\gamma}{\beta m^2}  \big\|\nabla_v \ln \psi \big\|^2
        \,\mathrm{d} t
%
    \end{aligned}
    \label{eq:underdamped_mart}
\end{equation}
Here, $n$ is the physical dimension of $v_t$, the first $n\gamma/m$
term results from $\tfrac{\gamma \Delta_v H}{m^2} =
\tfrac{\gamma\Delta_v \|v\|^{2}}{2m}$, and the second $n \gamma/m$
term derives from the $\gamma (\nabla_v\cdot v)/m$ term on the RHS of
Eq.~\eqref{psi_underdamped}. After cancelling terms and using
Eq.~\eqref{eq:underdamped} to express $\mathrm{d}v_{t}$ in terms of
$\mathrm{d}t$ and $\mathrm{d}B_t$, we find

\begin{equation}
    \begin{aligned}
        \mathrm{d} \theta_t = &  -\gamma \beta \|v\|^2 \,\mathrm{d}t +
       \sqrt{2 \gamma \beta }\, v\cdot \mathrm{d}B_t 
        - \mfrac{\gamma}{m}v\cdot
        \nabla_v \ln\psi \,\mathrm{d}t +
        \sqrt{\mfrac{2 \gamma}{\beta m^2}}\nabla_v \ln \psi 
        \, \mathrm{d}B_t
        - \mfrac{\gamma}{\beta m^2} \big\|\nabla_v \ln \psi \big\|^2 \,\mathrm{d}t \\
        = & - \mfrac{1}{2} \Big\|\sqrt{2 \gamma \beta } v 
        +\sqrt{\mfrac{2 \gamma}{\beta m^2}}\nabla_v
        \ln \psi \Big\|^2\mathrm{d}t +  \Big[\sqrt{2 \gamma \beta } v
          +\sqrt{\mfrac{2 \gamma}{\beta m^2}}\,\nabla_v
          \ln \psi \Big]\cdot \mathrm{d}B_t.
    \end{aligned}
    \label{eq:underdamped_mart2}
\end{equation}
In the last equality, we rearrange the terms to show that $\mathrm{d}
\theta_t = a_t \cdot \mathrm{d}B_t - \frac{1}{2} \|a_t\|^2
\,\mathrm{d}t$ for some process $a_t$ (in this case, $a_t\equiv
\sqrt{2 \gamma \beta } v_t + \sqrt{\tfrac{2 \gamma}{\beta
    m^2}}\nabla_v \ln \psi_t$). As with Eq.~\eqref{dtheta2}, this form
immediately implies that $\exp(\theta_t)$ is an exponential martingale:
\begin{equation}
    \begin{aligned}
      \mathrm{d} e^{\theta_t}  = & e^{\theta_t}\left[\mathrm{d}\theta_{t}+
        \mfrac{1}{2}\big(\mathrm{d}\theta_{t}\big)^{2}\right]\\
      = & e^{\theta_{t}}\left[a_t \cdot \mathrm{d}B_t
        -\mfrac{1}{2} \|a_t\|^2 \mathrm{d}t
        + \mfrac{1}{2} \|a_t\|^2 \mathrm{d}t \right] \\
      = & e^{\theta_t} a_t \cdot \mathrm{d}B_t.
    \end{aligned}
    \label{dtheta2}
\end{equation}
Consequently, provided that the Novikov condition holds, we find the
martingale property
\begin{equation}
    \mathbb{E} \left[ e^{-\beta (W_{t} - \Delta H_{t} + T \Delta
        \Sigma_{t})} \right] = 1.
    \label{eq:underdamped_martingale}
\end{equation}
Extension of Eq.~\eqref{eq:underdamped_martingale} to a bounded
stopping time $\tau$ follows immediately from the optional stopping
theorem for martingales.

\paragraph{Manzano's result for underdamped Langevin dynamics.}
In order to obtain the classical work theorem, we choose a specific
time $t_1$, and set $\psi(x,v,t_1) = \rho(x,v,t_1)$. It can be shown
that $\psi(-x,v, t_1 - t)$ follows a canonical Fokker-Planck equation;
thus $\int \psi(x,v,t) \mathrm{d}x \mathrm{d}v=1$.  Similarly, one can
derive
\begin{equation}
  \mathbb{E} \left[ e^{-\beta (W_{\tau} - \Delta F_{\tau}) - \delta_{\tau}} \right]
  = 1,\,\,  \forall ~\text{bounded stopping time}~\tau \leq t_1,
\end{equation}
where $\delta_{\tau} \equiv \ln \rho(x_{\tau},v_{\tau},\tau) - \ln
\psi(x_{\tau},v_{\tau},\tau)$ when $\psi$ is chosen such that
$\psi(x,v,t_1) = \rho(x,v,t_1)$.

\paragraph{Deterministic limits.} Our derivation also carries through
in the deterministic limits of zero-friction ($\gamma \to 0$) and
zero-temperature ($T\to 0$). In the zero-friction limit, the noise
term in Eq.~\eqref{eq:underdamped} vanishes and the acceleration
follows deterministic Hamiltonian dynamics \emph{without friction}.
We can directly compute the dimensionless entropy production of the
system: $\sigma_t \equiv \beta( H_t-W_t - T S_t) = \beta(H_t - W_t) +
\ln \rho(x_t,v_t,t)$. According to Liouville's theorem,
$\frac{\mathrm{d}}{\mathrm{d}t}\rho(x_t,v_t, t)=0$ in the
deterministic limit. Thus,
\begin{equation}
  \begin{aligned}
    \mathrm{d}\sigma_t = & \beta\mathrm{d}(H_t - W_t) \\
    \: = & \mfrac{\gamma}{m} \big(n
    - \beta \|v\|^{2} \big) \mathrm{d}t
    + \sqrt{2 \gamma\beta}\, v \cdot \mathrm{d} B_t,  
  \end{aligned}
\end{equation}
and in the $\gamma \to 0$ limit, we have
$\sigma_t = \sigma_0,\,\,\,  \text{almost surely}$.
%
In other words, when the entropy is evaluated by the information
entropy of the probability density, there is no entropy production in
the deterministic limit even for a non-autonomous non-conservative
Hamiltonian system.

In the zero-temperature limit ($\beta \rightarrow + \infty$), there is
no contribution from the entropy term to the free energy difference
and we have
\begin{equation*}
  \mathrm{d}(H_t - W_t) = - \frac{\gamma}{m} \|v\|^2 \mathrm{d}t,
\end{equation*}
which shows that energy is dissipated into the environment at rate
$\gamma \|v\|^2/m$.

\paragraph{Fluctuation-dissipation relation -- underdamped limit.} 
Similar to the overdamped case, we can explicitly compute $\mathrm{d}e^{\theta_t}$
in cases where the fluctuation-dissipation relation is not assumed.
The general underdamped Langevin equation is given by
\begin{equation}
  \begin{aligned}
  \mathrm{d}x_t &= v_t\,\mathrm{d}t,\\
  m\,\mathrm{d}v_t &= \Big[-\gamma\,v_t -\nabla_xU(x_t,t)+f(x_t,t)\Big]\,
  \mathrm{d}t  + \gamma\sqrt{2D}\,\mathrm{d}B_t,
  \end{aligned}
\end{equation}
and the corresponding backward equation for $\psi$ is given by
\begin{equation}
  -\partial_t \psi(x, v, t)=v \cdot \nabla_x \psi+\mfrac{\gamma^2 D}{m^2} \Delta_v \psi
  -\mfrac{1}{m} \nabla_v \cdot\Big[\big(-\gamma v+\nabla_x U-f\big) \psi\Big].
\end{equation}
Following the same procedure as in Eqs.~\eqref{eq:underdamped}--\eqref{dtheta2} 
and detailed in Appendix~\ref{sec:underdamped}, 
we can find 
\begin{equation}
  \begin{aligned}
    \mathrm{d}e^{\theta_t} = e^{\theta_t}\bigg[
      \Big(D-\mfrac{1}{\beta\gamma}\Big) \mfrac{\gamma^2 \beta}{m}
    \Big( \beta m \|v\|^2 + 2v_t\cdot\nabla_v\ln\psi _t
      + n\Big)\,\mathrm{d}t + \gamma\sqrt{2D}\Big(\beta\,v
      +\mfrac{1}{m}\nabla_v\ln\psi\Big)\cdot \mathrm{d}B_t 
    \bigg].
  \end{aligned}
\end{equation}

When $D \neq 1/(\beta \gamma)$, the drift term is typically
nonzero as the factor $\beta m \|v\|^2$ depends on the current velocity
$v$ of the particle. In general, $\mathbb{E} \left[\beta m \|v\|^2 +
2v_t\cdot\nabla_v\ln\psi _t + n\right]$ measures the deviation of
velocity distribution from the Maxwell-Boltzmann distribution where 
$v \propto \exp (-\frac{\beta m}{2} \|v\|^2)$. When $v$ follows 
the Maxwell-Boltzmann distribution, $\mathbb{E}  \left[\beta m \|v\|^2 +
2v_t\cdot\nabla_v\ln\psi _t + n\right] = 0$.

\section{Numerical examples}
\paragraph{Drift-diffusion process.}
We first provide a simple but useful numerical example on which to
apply our work theorem. Our example is inspired by the continuous-time
version of the kinetic proofreading (KPR) mechanism
\cite{Hopfield1974oct}.  The KPR mechanism is a nonequilibrium process
is typically invoked in DNA replication or cell signaling that
amplifies differences in the unbinding rates of different ligands in
order to increase the specificity of ligand recognition. The original
kinetic proofreading mechanism relies on multiple discrete activation
steps on the ligand-receptor complex. In our previous work
\cite{li2024reliable}, we generalized the discrete activation process
to a continuum process in the limit of large number of activation
steps. In this continuum limit, can a work theorem can provide a
fundamental bound on the energy-information trade-off?
\added{Answering this question can provide additional geometric
  insight into the speed-energy-accuracy trade-off in nonequilibrium
  systems \cite{Jrmie2024Feb}.}

\begin{figure}
  \centering
  \includegraphics[width=4.5in]{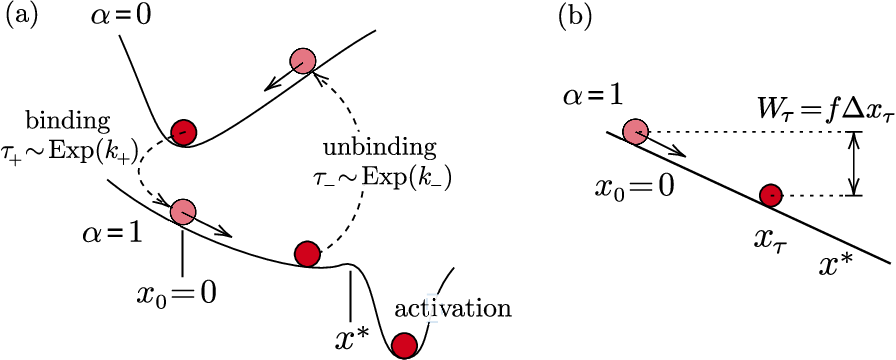}
  \caption{(a) Schematic representation of the continuum kinetic
    proofreading process. The receptor state is defined by the
    activation level $x$ and the ligand-binding status $\alpha \in
    \{0,1\}$. When the ligand is absent ($\alpha = 0$), the system
    relaxes to the stable state at $x = 0$. Ligand binding ($\alpha =
    1$) introduces an external force $f$ that biases the activation
    dynamics, enabling transitions to higher activation
    levels. Binding and unbinding events follow exponential waiting
    time distributions, $\tau_+ \sim \operatorname{Exp}(k_+)$ and
    $\tau_- \sim \operatorname{Exp}(k_-)$, respectively. Activation
    occurs when $x > x^{*}$, with the probability of activation
      depending exponentially on the unbinding rate $k_-$,
    facilitating high selectivity in ligand recognition. Ligand
    unbinding before ($\alpha = 0$) resets the activation level to $x
    = 0$, analogous to erasing memory as constrained by Landauer's
    principle.  (b) A Simplified drift-diffusion representation of the
    process with bound ligand ($\alpha =1$). The dynamics reduce to an
    overdamped Langevin equation under a constant force $f$.  Starting
    from $x_0 = 0$, the system evolves according to $\gamma \mathrm{d}
    x_t = f \mathrm{d} t + \sqrt{\frac{2 \gamma}{\beta}} \mathrm{d}
    B_t$. The work performed by the force is given by $W = f \Delta
    x_\tau$, where $\Delta x_\tau$ is the displacement over time
    $\tau$. This model captures the fundamental principles of energy
    expenditure and memory erasure in ligand-receptor systems.}
  \label{fig:kpr_schematic}
\end{figure}

A simple schematic of a continuum kinetic proofreading process is
shown in Fig.~\ref{fig:kpr_schematic}(a). Let $(x,\alpha) \in
\mathbb{R} \times \{0,1\}$ be the state of the receptor. $x$
represents the level of activation or ``reaction coordinate'' and
$\alpha=1$ indicates the presence of ligand. Without ligand
($\alpha=0$), the Hamiltonian $H(x)$ is minimized at $x=0$. With bound
ligand ($\alpha=1$), an external force $f$ arises such that
$-\partial_x H(x)+f > 0$.  Ligand-receptor binding and unbinding are
assumed independent of the activation level $x$ and follow exponential
waiting time distributions, $\tau_+ \sim \operatorname{Exp}(k_+)$ and
$\tau_- \sim \operatorname{Exp}(k_-)$, respectively. The complex is
considered activated if $x > x^{*}$ for some threshold $x^{*}$. During the
activation process ($\delta = 1$), the activation probability $P(x >
A)$ is roughly exponentially dependent on the unbinding rate $k_-$,
thus enabling highly selective ligand recognition. The deactivation
process ($\alpha = 0$) can be considered as erasure of memory (i.e.,
the activation level is reset to $x=0$). Previous application of the
work theorem leads to the Landauer principle, which suggests that
erasing one bit of information requires $k_{\rm B}T \ln 2$ work.

As a proof of concept, we further simplify the dynamics of 
the continuum proofreading process to a drift-diffusion process,
as shown in Fig.~\ref{fig:kpr_schematic}(b). In the
drift-diffusion process, we combine the Hamiltonian and force
into a constant total force $f$ so that Eq.~\eqref{eq:overdamped}
can be simplified to
\begin{equation}
    \gamma \mathrm{d} x_t = f ~\mathrm{d} t + \sqrt{
      \mfrac{2 \gamma}{\beta}}\, \mathrm{d} B_t.
      \label{eq:example}
\end{equation}
Given an initial position (state) $x_0 = 0$,
the solution to Eq.~\eqref{eq:example} is
\begin{equation}
    x_t = \mfrac{f}{\gamma} t + 
    \sqrt{\mfrac{2 \gamma}{\beta}} B_t.
\end{equation}
The backward process $\psi(x,t)$ given by Eq.~\eqref{eq:backward}
now becomes
\begin{equation}
  -\partial_t \psi = - 
  \mfrac{f}{\gamma} \partial_x \psi
  + \mfrac{1}{\beta \gamma} \partial_x^2 \psi.
  \label{eq:backward_example}
\end{equation}
Using an initial $\delta$-function distribution at $x=0$, we can
construct a family of solutions to Eq.~\eqref{eq:backward_example} by
time-reversing the solution of the corresponding Fokker-Planck
equation using different terminal times $t_2$. Specifically,
\begin{equation}
  \begin{aligned}
    p(x, t) & =\sqrt{\mfrac{\beta \gamma}{4\pi t}} 
    \exp \left[
      -\mfrac{\beta \gamma \left(x-x_0-\tfrac{f}{\gamma} t\right)^2}
        {4 t}
        \right],\\
    \psi(x,t \mid t_2) & = p(x, t_2 -t), \quad \text{for } t < t_2.
  \end{aligned}
\end{equation}
In general, $\psi(x,t \mid t_2)$ solves Eq.~\eqref{eq:backward_example}
up to time $t_2$. 

\begin{figure}
  \centering
  \includegraphics[width=4in]{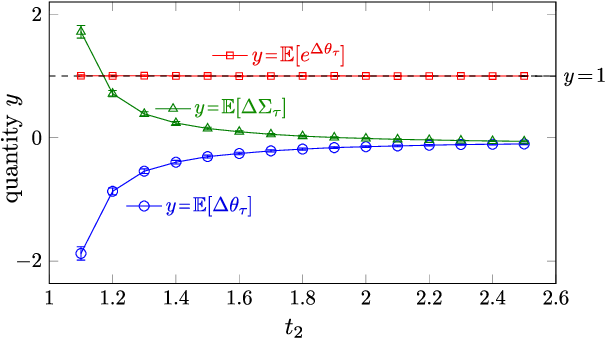}
  \caption{Numerical results for the quantities
    $\mathbb{E}[e^{\theta_{\tau}}]$ (red), $\mathbb{E}[\theta_{\tau}]$
    (blue), and $\mathbb{E}[\Delta \Sigma_{\tau}]$ (green) for
    different choices of $t_2$. The red line is at unity, consistent
    with our work theorem. By contrast, $\mathbb{E}[\theta_{\tau}]$
    and $\mathbb{E}[\Delta \Sigma_{\tau}]$ vary with $t_2$, shown by
    the blue and green curves, respectively.
    \label{fig:numerical_example_results}}
\end{figure}

Now, consider the exponentially distributed stopping time $\tau_{-}
\sim \operatorname{Exp}(k_{-})$ for the unbinding process.  In order
to apply the optional sampling theorem, we bound the stopping time by
the desired terminal time $t_1$: $\tau = \min \{ \tau_-, t_1 \}$.
Then, for concreteness, we choose $f = 1$, $\gamma = 1$, and $k_-=1$ and
numerically evaluate the expectations of normalized ``entropy''
production $\theta_{\tau}$, ``entropy'' change $\Delta
\Sigma_{\tau}$, and $e^{\theta_{\tau}}$ at the stopping time
$\tau$ for different choices of $\psi(x,t \mid t_2)$.  The results are
plotted in Fig.~\ref{fig:numerical_example_results} and are consistent
with predictions of our work theorem; the red curve in
Fig.~\ref{fig:numerical_example_results} shows that
$\mathbb{E}[e^{\theta_\tau}]=1$ for all choices of $t_2$. By
contrast, $\mathbb{E}[\theta_{\tau}]$ (blue curve) and
$\mathbb{E}[\Delta \Sigma_{\tau}]$ (green curve) vary with $t_2$.  It is
noteworthy that this numerical example does not admit a classical work
theorem counterpart, as the initial distribution of $x_0$ is not a
density function and its trajectorywise entropy is not well-defined.

This calculation is helpful for understanding how the 
``entropy'' $\Sigma$ changes with time for different 
choices of the backward process $\psi(x,t \mid t_2)$. 
In order to better understand the energy-information 
trade-off in the continuum proofreading mechanism, 
we need to relate the backward-process associated entropy
$\Sigma(x, \tau \mid t_2)$  to the actual entropy 
$- k_B \ln \rho(x_{\tau})$ at the stopping time $\tau$.
We left this as future work.

\paragraph{Violation of the fluctuation-dissipation relation
  in the Ornstein-Uhlenbeck process.} We now consider the
Ornstein-Uhlenbeck (OU) process defined by 
\begin{equation}
  \mathrm{d} x_t = - \mfrac{k}{\gamma} x_t \mathrm{d}t + \sqrt{2D}\,
  \mathrm{d}B_t,
  \label{eq:ou}
\end{equation}
which is Eq.~\eqref{eq:overdamped_general} with $U(x) = \frac{k}{2}
x^2$ and $f(x) =0$. We choose a Gaussian initial distribution
$\rho(x,0) = \mfrac{1}{\sqrt{2\pi} \sigma} e^{-
  \frac{x^2}{2\sigma^2}}$ with variance $\sigma^2$ such that the
classical work theorem can be applied to this scenario.

The probability density $\rho(x,t)$ of the OU process is explicitly
given by
\begin{equation}
\rho(x,t) =  \frac{1}{\sqrt{2\pi}\, \sigma(t)} e^{- \frac{x^2}{2\sigma^2(t)}},\quad 
\sigma^2(t) = \sigma^2 e^{-kt/\gamma} + \mfrac{D \gamma}{k}\big(1-e^{-kt/\gamma}\big).
\end{equation}
To demonstrate the general applicability of our
forward-backward-decoupled work theorem, we choose $\psi(x,t)$ to be
the stationary distribution of the OU process:
\begin{equation}
\psi(x,t) = \sqrt{\mfrac{k}{2\pi D \gamma}}\, e^{- \frac{k x^2}{2D \gamma}}.
\end{equation}

We again conduct numerical simulations of Eq.~\eqref{eq:ou} with
$\gamma = 1$, $k = 1$, and varying diffusion coefficient $D$ up to a
fixed time $t_1 = 2$. For each $D$, we simulate $10^7$ trajectories to
ensure convergence of the expectation. We evaluate both the classical
exponentiated entropy production $\mathbb{E} \left[e^{\beta (W_t -
    \Delta F_t)}\right]$ and the generalized exponentiated entropy
production $\mathbb{E} \left[e^{\theta_t}\right]$ at time $t_1$. Note
that the exact expressions of $\rho(x,t)$ and $\psi(x,t)$ are used to
compute the entropies in the two cases, respectively.

\begin{figure}
  \centering
  \includegraphics[width=3.5in]{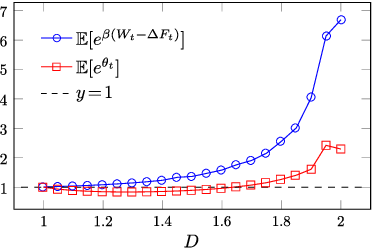}
  \caption{Numerical results of the exponentiated entropy production
    $\mathbb{E} \left[e^{\beta (W_t - \Delta F_t)}\right]$ (blue) and
    $\mathbb{E} \left[e^{\theta_t}\right]$ (red) for the OU process
    with different diffusion coefficients $D$. In both cases, when
    $D=1/(\beta \gamma)=1$, both quantities are equal to unity.
    \label{fig:ou_results}}
\end{figure}

As is shown in Fig.~\ref{fig:ou_results}, when $D=1/(\beta \gamma)=1$,
both $\mathbb{E} \left[e^{\beta (W_t - \Delta F_t)}\right]$ and
$\mathbb{E} \left[e^{\theta_t}\right]$ are equal to unity, consistent
with the classical and generalized work theorems, respectively.  As
$D$ deviates from $1/(\beta \gamma)=1$, $\mathbb{E} \left[e^{\beta (W_t
    - \Delta F_t)}\right]$ and $\mathbb{E} \left[e^{\theta_t}\right]$
can be either greater or less than unity, indicating a violation of
the work theorem when the fluctuation-dissipation relation is not
satisfied.

\section{Discussion and Conclusions}
Work theorems are fundamental in that they characterize the property
of entropy production in stochastic systems and generalize the second
law of thermodynamics to nonequilibrium systems. In this work, we have
shown that the work theorem can be further generalized to a broader
class of stochastic systems by introducing a specific backward process
that is essentially a time-reversed diffusion process with
time-reversed Hamiltonian and force. However, unlike the path-integral
formulation of Crooks' theorem, the backward process does not
necessarily share a common distribution with the original ``forward''
process at any given time $t$. Specifically, when the Hamiltonian and
force are time-independent, we provide a new equality that relates the
work and energy change of the system to the stationary entropy at
steady state.

A key insight of our work is that forward and backward processes can
be decoupled as shown in Fig.~\ref{fig:schematic2}(a). The generalized
work theorem also \textit{decouples} the initial sample $x_0$ from the
initial distribution $\rho(x,0)$, as shown in
Fig.~\ref{fig:schematic2}(b). To be more specific, we have shown that
$\mathbb{E}\big[e^{\Delta \theta_\tau} \mid x_0\big] = 1$ for any
bounded stopping time $\tau$ and any initial sample $x_0$ in
Eq.~\eqref{eq:main}. The usual formulation of the work theorem
involves an expectation or integral over the initial distribution
$\rho(x,0)$, \textit{i.e.}, $\int \mathbb{E} \big[e^{\Delta \sigma_t}
  \mid x_0\big] \rho(x_0,0) \,\mathrm{d} x_0=1$. Our analysis provides
a stronger result: the work theorem holds for any initial sample $x_0$
regardless of the initial distribution. It also suggests new
directions of investigation such as the physical interpretation of the
backward process with different ``initial'' conditions. It is clear
that when the backward process can be interpreted as a time-reversed
process (with proper reflection in the case of underdamped dynamics),
its probability distribution coincides with that of the forward
process at a specific time. However, the martingale property and
Eq.~\eqref{eq:main} do allow for different choices of the initial
condition of the backward process. \added{The simple stationary form
  of the generalized work theorem given in Eq.~\eqref{eq:stationary}
  is also of particular interest. Since the entropy term is
  time-independent, we do not need to compute the time evolution of
  $\rho(x,t)$, allowing easier analytical or numerical application of
  the generalized work theorem to complex systems.}

We have also extended our methods to underdamped Langevin dynamics
which is of particular interest in view of its natural connection to
the deterministic Hamiltonian dynamics in the limit of vanishing
friction. While the underdamped Langevin dynamics is widely
anticipated to follow the same work theorem as the overdamped Langevin
dynamics, our work provides a formal proof of this.
\deleted{The underdamped
Langevin dynamics is of particular interest in view of its natural
connection to the deterministic Hamiltonian dynamics in the limit of
vanishing friction.}
Nonequilibrium thermodynamics laws that are applicable to both
underdamped and overdamped Langevin dynamics have been of recent
interest \cite{dago2022dynamics,Lyu2024December}.  When measured by
the information entropy of the probability density, the entropy
production vanishes in the deterministic limit, even for a
non-autonomous non-conservative Hamiltonian system. This result is
expected but highlights the importance of understanding the definition
of entropy in macroscopic systems, which may require coarse-graining
measurement of the phase space
\cite{barkan2024convergencephasespacedistributions}. On the other
hand, because of the introduction of velocity in the underdamped
dynamics, it is possible to separate the internal ``temperature'' of
the system (defined by the kinetic energy of the system) from the
external temperature (of the heat bath)
\cite{dago2022dynamics}. Consequences of this separation are yet to be
explored. Coupled thermal machines operating at different temperatures
\cite{Leighton2024November} is also another promising direction in
which to extend work theorems.

Our It\^o calculus-based derivation enables us to analyze the work
theorem in cases where the fluctuation-dissipation relation is not
satisfied. Our theoretical and numerical results emphasize that the
work theorems require careful adaptation for nonequilibrium
systems. This highlights how \added{violation of the
  fluctuation-dissipation relation}, common in active matter and
biological systems, generates intrinsic entropy production that cannot
be reconciled with equilibrium-based theorems.

Different choices of the backward process may shed light on important
nonequilibrium processes. For example, from the original work theorem,
one can derive the Landauer principle by setting $\Delta H=0$, which
states that the work required to erase one bit of information is
$k_{\rm B}T \ln 2$; this has been experimentally verified and
extended to include the effects of erasure speed
\cite{proesmans2020finite,dago2022dynamics}. Faster erasure requires
more total work. The original work theorem only provides the original
bound of the Landauer principle. It would be worthwhile to investigate
whether the generalized work theorem can incorporate the effects of
erasure speed and provide a more accurate bound for the work required
to erase information.

\deleted{The simple stationary form of the generalized work theorem in
  Eq.~\eqref{eq:stationary} is also of particular interest. Since the
  entropy term is time-independent, the generalized work theorem can
  be more easily analytically or numerically applied to complex
  systems. Cells and other living systems are known to consume energy
  to extract more information from the environment; the generalized
  work theorem may also provide bounds to the energy-information
  trade-off in these systems. From a thermodynamic point of view, it
  could complement existing understanding of molecular machines in the
  cell \cite{Brown2019August}. Recently, a form of fluctuation theorem
  has been developed \cite{Piephoff2025Jan} to detect breaking of
  detailed balance in the biomolecular networks. Traditionally, the
  work theorem is usually a consequence of the corresponding
  fluctuation theorem. It would be interesting to see if the
  generalized work theorem developed in this paper admits a
  corresponding fluctuation theorem.}

Lastly, from both theoretical and applied perspectives, one
inconvenient aspect of the work theorem is that it requires working in
unbounded spaces and the Novikov condition to hold. Additionally,
experimental measurement may introduce discretization of the continuum
distribution of states.  Systems such as stochastic chemical reactions
and other biological processes are more realistically described by
bounded and/or discrete spaces.  This has been numerically explored by
Manzano \textit{et al}. \cite{Manzano2021February} using a simple
two-state model. \added{However, the theoretical understanding of the
  work theorem in general discrete spaces such as chemical reaction
  networks \cite{Schmiedl2007January} remains relatively unexplored.
  Doing so could require using additional mathematical tools such as
  the cycle representation theory of Markov chains
  \cite{Jiang2023March} and invoking insights such as the generalized
  Legendre-Frenchel transform and response theory for the steady state
  of nonequilibrium Markov chains
  \cite{Yang2024OctA,Yang2024OctA,Zheng2024Mar}.}

\vspace{5mm}
\subsection*{Data Availability} There were no data generated or
analyzed in this study.

\section*{Declarations}

\subsection*{Conflict of interest} The authors have no conflicts of
interest relevant to the content of this article.


\newpage
\renewcommand{\thefigure}{A\arabic{figure}}
\renewcommand{\theequation}{A\arabic{equation}} \setcounter{figure}{0}
\setcounter{equation}{0}

\appendix

\section{Mathematical Appendix}

\subsection{Overdamped Langevin dynamics with $D \neq 1/(\beta \gamma)$}
\label{app:general_derivation}

In the main text, the derivation of $\mathrm{d}\theta_t$ and the
martingale property of $\exp(\theta_t)$ relies on the specific choice
of the diffusion coefficient $D = 1/(\beta \gamma)$. Here, we
present a more general derivation that avoids this assumption.
Instead, we consider the stochastic dynamics
\begin{equation}
  \mathrm{d}x_t = \mfrac{1}{\gamma}
  \big[-\nabla H(x_t,t) + f(x_t,t)\big]\,\mathrm{d}t 
    + \sqrt{2D}\,\mathrm{d}B_t,
\end{equation}
where $D$ is a general diffusion coefficient that is no longer tied to
the fluctuation-dissipation relation $D=1/(\beta \gamma)$.

\paragraph*{Setup}

We define the backward process by its probability density
$\psi(x,t)$ that satisfies
\begin{equation}
    -\partial_t \psi(x,t) 
    = \mfrac{1}{\gamma} \nabla \cdot \Big[ \big(\nabla H(x,t)
      - f(x,t)\big) \psi(x,t) \Big] 
    + D \Delta \psi(x,t).
    \label{eq:backward_general}
\end{equation}
We also recall the definitions:
\[
    W_t = \int_0^t f(x_s,s) \circ \mathrm{d}x_s 
    + \int_0^t \partial_t H(x_s,s)\,\mathrm{d}s,
\]
and
\[
    \Sigma(x_t,t) = -k_B \ln \psi(x_t,t).
    \]
and consider Eq.~\ref{THETA_SIGMA}
\begin{equation}
    \theta_t = - \beta W_t + \beta H(x_t,t) - \beta T \Sigma(x_t,t)
    \label{eq:theta_def_general}
\end{equation}
but with $\psi(x_t,t)$ in the definition of $\Sigma(x_t,t)$ determined
by Eq.~\eqref{eq:backward_general} in which $D$ is thus far arbitrary.

\paragraph*{Derivation of \texorpdfstring{$\mathrm{d}\theta_t$}{d theta\_t}}

First, we compute $\mathrm{d}H(x_t,t)$ using Ito's lemma:
\begin{equation}
\begin{aligned}
    \mathrm{d}H & = \partial_t H\,\mathrm{d}t 
    + \nabla H \cdot \mathrm{d}x_t 
    + \mfrac{1}{2}\Delta H (\mathrm{d}x_t)^2.
\end{aligned}
\end{equation}
Since $(\mathrm{d}x_t)^2 = 2D\,\mathrm{d}t$, we find

\begin{equation}
\mathrm{d}H = \partial_t H\,\mathrm{d}t 
    + \nabla H \cdot \mathrm{d}x_t 
    + D \Delta H\,\mathrm{d}t.
    \label{dH}
\end{equation}
Then, from the definition of $W_t$, we have
\begin{equation}
  \mathrm{d}W_t = f(x_t,t) \circ \mathrm{d}x_t
  + \partial_t H(x_t,t)\,\mathrm{d}t.
\end{equation}
Converting this Stratonovich integral to an Ito integral, we find
\begin{equation}
    f(x_t,t) \circ \mathrm{d}x_t = f(x_t,t)\cdot\mathrm{d}x_t 
    + D \nabla \cdot f(x_t,t)\,\mathrm{d}t
\end{equation}
and
\begin{equation}
    \mathrm{d}W_t = f \cdot \mathrm{d}x_t 
    + \partial_t H\,\mathrm{d}t 
    + D (\nabla \cdot f)\,\mathrm{d}t.
    \label{dW}
\end{equation}
Finally, we decompose $\mathrm{d}\Sigma$ as 
\begin{equation}
    \mathrm{d}\Sigma =  -k_{\rm B}\, \mathrm{d}\ln \psi(x_{t},t) =
    -k_{\rm B} \bigg(\frac{\mathrm{d}\psi}{\psi}-\mfrac{1}{2}
    \Big[\frac{\mathrm{d}\psi}{\psi}\Big]^{2}\bigg).
  \label{dSigma}
\end{equation}
Upon using Eqs.~\eqref{dH}, \eqref{dW}, and \eqref{dSigma}
in Eq.~\eqref{eq:theta_def_general}, we find
%
%
\begin{equation}
\begin{aligned}
  \mathrm{d}\theta_t = & -\beta \mathrm{d}W_t + \beta \mathrm{d}H
  - \beta \mathrm{d}\Sigma \\
  = & -\beta\big(f\cdot\mathrm{d}x_t + \partial_t H\,\mathrm{d}t
  + D \nabla\cdot f\,\mathrm{d}t\big) + \beta\big(\partial_t H\,\mathrm{d}t
  + \nabla H\cdot\mathrm{d}x_t + D\Delta H\,\mathrm{d}t\big)
  + \frac{\mathrm{d}\psi}{\psi} - \frac{[\mathrm{d}\psi]^2}{2\psi^2}.
\end{aligned}
\label{dtheta_psi}
\end{equation}
We now decompose the $\mathrm{d}\psi/\psi$ and $(\mathrm{d}\psi/\psi)^{2}$
by applying Ito's lemma to $\psi$,

\begin{equation}
  \mathrm{d}\psi = \partial_{t}\psi\,\mathrm{d}t
  +\nabla\psi\cdot\mathrm{d}x_{t} + \mfrac{1}{2} \Delta \psi
  (\mathrm{d}x_{t})^{2}, 
\end{equation}
using Eq.~\eqref{eq:backward_general} to eliminate $\partial_{t}\psi$, and
substituting $(\mathrm{d}x_t)^2 = 2D\,\mathrm{d}t$, to find

\begin{equation}
\begin{aligned}
  \mathrm{d}\psi  = & -\mfrac{1}{\gamma}\nabla\cdot\big[(\nabla H - f)\psi\big]
  \,\mathrm{d}t +  \nabla \psi \cdot \mathrm{d}x_t.
\end{aligned}
\end{equation}
Using this to construct $\mathrm{d}\psi/\psi$ and
$(\mathrm{d}\psi/\psi)^{2}$, we find

\begin{equation}
  \begin{aligned}
 \frac{\mathrm{d}\psi}{\psi} = & -\mfrac{1}{\gamma}(\Delta H -
 \nabla\cdot f)\,\mathrm{d}t -\mfrac{1}{\gamma}(\nabla H -
 f)\cdot\nabla\ln\psi\,\mathrm{d}t + \nabla\ln\psi \cdot \mathrm{d}x_t
 \\ \frac{[\mathrm{d}\psi]^2}{2\psi^2} = & 
 \frac{(\nabla\psi\cdot\mathrm{d}x_t)^{2}}{2\psi^{2}} =
 D\|\nabla\ln\psi\|^2\,\mathrm{d}t.
  \end{aligned}
  \label{ddpsi}
\end{equation}

\noindent Substitution of Eqs.~\eqref{ddpsi} into
Eq.~\eqref{dtheta_psi} leads to
  
\begin{equation}
  \begin{aligned}
\mathrm{d}\theta_t 
=  & \left[\big(\beta D -\tfrac{1}{\gamma}\big) \big(\Delta H
  - \nabla\cdot f\big) - \mfrac{1}{\gamma}(\nabla H -f)\cdot\nabla\ln\psi 
  - D\big\|\nabla\ln\psi \big\|^2\right]\mathrm{d}t \\
\: & \qquad \quad + \big(\beta(\nabla H -f) + \nabla \ln\psi\big)\cdot \mathrm{d}x_t
  \end{aligned}
\label{eq:theta_dx_dt}
\end{equation}
Since the forward SDE is $\mathrm{d}x_t = -\tfrac{1}{\gamma}(\nabla
H - f)\,\mathrm{d}t + \sqrt{2D}\,\mathrm{d}B_t$ we can replace
$\mathrm{d}x_t$ in Eq.~\eqref{eq:theta_dx_dt} to find
\begin{equation}
  \begin{aligned}
    \mathrm{d}\theta_t = & \Big(D - \mfrac{1}{\beta \gamma}\Big)\Big(
      \beta\big(\Delta H - \nabla\cdot f\big)-
      \big\|\nabla \ln \psi\big\|^2\Big) \mathrm{d}t
      - \mfrac{1}{\beta \gamma} \big\|\beta(\nabla H - f) +
      \nabla \ln \psi\big\|^2 \mathrm{d}t \\
      \: &  \qquad \quad + \sqrt{2D}\Big(\beta(\nabla H -f)
    + \nabla\ln\psi\Big)\cdot \,\mathrm{d}B_t.
  \end{aligned}
  \label{eq:theta_final}
\end{equation}
When $D = 1/(\beta \gamma)$, the last term in
Eq.~\eqref{eq:theta_final} vanishes, rendering it equivalent to
Eq.~\eqref{eq:theta_overdamped_final}.  In general,
\begin{equation}
\begin{aligned}
  \mathrm{d} e^{\theta_t} = e^{\theta_t} \Bigg[ \Big( D - \mfrac{1}{\beta \gamma} \Big)
    \Big( \beta^2 & \big\|\nabla H - f\big\|^2 + 2 \beta (\nabla H - f) \cdot \nabla \ln \psi
    + \beta (\Delta H - \nabla \cdot f) \Big) \mathrm{d}t \\[-4pt]
    \: &  + \sqrt{2D} \big(
    \beta (\nabla H - f) + \nabla \ln \psi \big) \cdot \mathrm{d}B_t\Bigg],
\end{aligned}
\end{equation}
from which it is apparent that $\mathrm{d}e^{\theta_t}$ is subject to
non-zero drift and $e^{\theta_t}$ is not a martingale in general.


\subsection{Underdamped Langevin dynamics with $D \neq 1/(\beta \gamma)$}
\label{sec:underdamped}
We now outline how dynamics that deviate from the
fluctuation-dissipation relation modify the work theorem.  Consider an
$n$-dimensional underdamped Langevin system described by

\begin{equation}
\begin{aligned}
\mathrm{d}x_t &= v_t\,\mathrm{d}t,\\
m\,\mathrm{d}v_t &= \Big[-\gamma\,v_t -\nabla_xU(x_t,t)+f(x_t,t)\Big]\,
\mathrm{d}t  + \gamma\sqrt{2D}\,\mathrm{d}B_t,
\end{aligned}
\label{eq:underdamped_nonFDT}
\end{equation}
where $x_t,\,v_t\in \mathbb{R}^n$, $m$, $\gamma$ and $D$ are
constants, and $B_t$ is a standard diagonal, $n$-dimensional Brownian
motion.  Let the backward process $\psi(x,v,t)$ be defined by the PDE
\begin{equation}
  -\partial_t \psi(x, v, t)=v \cdot \nabla_x \psi+\mfrac{\gamma^2
    D}{m^2} \Delta_v \psi-\mfrac{1}{m} \nabla_v
  \cdot\big[\left(-\gamma v+\nabla_x U-f\right) \psi\big].
\end{equation}
Other quantities are defined in the same way as in the main
text. 

To compute $\mathrm{d}e^{\theta_t}$ first apply It\^o formula to
$\mathrm{d}H_t$ to obtain

\begin{equation}
\mathrm{d}H_t= \Big[-\gamma\|v\|^2 + v\cdot f 
                    + \partial_tU +\mfrac{\gamma^2D\,n}{m}\Big]\,\mathrm{d}t 
            + \gamma\sqrt{2D}\,v\cdot\mathrm{d}B_t.
\label{eq:dH}
\end{equation}
From the definition of work and $\mathrm{d}x=v \mathrm{d}t$, we have

\begin{equation}
\mathrm{d}W_t = \Big[f(x_t,t)\cdot v_t + \partial_t U(x_t,t)\Big]\, \mathrm{d}t.
\label{eq:dW}
\end{equation}
Similarly, applying It\^o's formula to $\mathrm{d}\psi$, inserting the PDE
for $\psi$ to eliminate $\partial_t\psi$, and assuming that $f$ and $U$
are $v$-independent, we obtain

\begin{equation}
  \mathrm{d}\ln\psi = \left[-\mfrac{2\gamma}{m}\,v\cdot\nabla_v\ln\psi
    - \frac{\gamma n}{m} - \mfrac{\gamma^2D}{m^2}\big\|\nabla_v\ln\psi\big\|^2
    \right]\mathrm{d}t +
  \mfrac{\gamma\sqrt{2D}}{m}\,\nabla_v\ln\psi\cdot\mathrm{d}B_t.
\label{eq:dlnpsi}
\end{equation}

Given the above results, we now derive the differential of $\theta_t =
\beta (\Delta H_t - W_{t}) + \Delta \ln \psi_t$:

\begin{equation}
  \mathrm{d}\theta_t = \left[-\beta\gamma\|v\|^2
    + \beta\mfrac{\gamma^2D\,n}{m} -
    \mfrac{2\gamma}{m}\,v\cdot\nabla_v\ln\psi
    - \mfrac{\gamma n}{m} - \mfrac{\gamma^2D}{m^2}
    \big\|\nabla_v\ln\psi\big\|^2\right]\,\mathrm{d}t
  + \gamma\sqrt{2D}\left[\beta\,v +
    \mfrac{1}{m}\nabla_v\ln\psi\right]\cdot \mathrm{d}B_t.
\label{eq:dtheta}
\end{equation}
After expanding and collecting terms, $\mathrm{d}e^{\theta_t}$ becomes

\begin{equation}
\begin{aligned}
  \mathrm{d}e^{\theta_t} = e^{\theta_t}\bigg[\Big(D-\mfrac{1}{\beta\gamma}\Big)
    \mfrac{\gamma^2 \beta}{m}
  \Big( \beta m \|v\|^2 + 2v_t\cdot\nabla_v\ln\psi _t
    + n\Big)\,\mathrm{d}t + \gamma\sqrt{2D}\Big(\beta\,v
    +\mfrac{1}{m}\nabla_v\ln\psi\Big)\cdot \mathrm{d}B_t \bigg].
\end{aligned}
\label{eq:final}
\end{equation}
Notice that $D= 1/(\beta \gamma)$ is a sufficient condition for
the drift term to vanish and our generalized work theorem to hold. The
term $\beta m\|v\|^2+2 v \cdot \nabla_v \ln \psi+n$ can be a measure
of the violation of the the equipartition theorem.

If the system is in equilibrium such that $\psi(v) \propto e^{-
  \frac{\beta m}{2} \|v\|^2}$ and the equipartition theorem holds
($\mathbb{E}[\beta m \|v\|^2]=n$), then on average

\begin{equation}
  \mathbb{E}\left[\beta m \|v\|^2 + 2v\cdot\nabla_v\ln\psi + n\right]
  = n - 2n + n = 0.
\end{equation}
In this case, the generalized work theorem holds for underdamped
Langevin dynamics even when $D \neq 1/(\beta \gamma)$ provided
$\psi(v,t) \propto e^{- \frac{\beta m}{2} \|v\|^2}$ for all $t$.

\newpage 
\bibliographystyle{unsrt}

\bibliography{work_theorem}
\end{document}